\documentclass[a4paper, 10pt, conference]{ieeeconf}      

\IEEEoverridecommandlockouts                              
\usepackage{graphicx}
\usepackage{amsmath}
\usepackage{balance}
\usepackage{algorithm}
\usepackage[noend]{algpseudocode}
\usepackage[dvipsnames]{xcolor}

\title{Valuating User Data in a Human-Centric Data Economy}

\author{ \parbox{3 in}{\centering Marius Paraschiv \\
        IMDEA Networks\\
        Legan\'{e}s -- Madrid\\
        {\tt\small marius.paraschiv@imdea.org}}
        \hspace*{ 0.5 in}
        \parbox{3 in}{ \centering Nikolaos Laoutaris \\
        IMDEA Networks \\
        Legan\'{e}s -- Madrid\\
        {\tt\small nikolaos.laoutaris@imdea.org}}
}

\begin{document}
\maketitle
\pagestyle{plain}

\begin{abstract}
The idea of paying people for their data is increasingly seen as a promising direction for resolving privacy debates, improving the quality of online data, and even offering an alternative to labour-based compensation in a future dominated by automation and self-operating machines. In this paper we demonstrate how a Human-Centric Data Economy would compensate the users of an online streaming service. We borrow the notion of the Shapley value from cooperative game theory to define what a fair compensation for each user should be for movie scores offered to the recommender system of the service. Since determining the Shapley value exactly is computationally inefficient in the general case, we derive faster alternatives using clustering, dimensionality reduction, and partial information. We apply our algorithms to a movie recommendation data set and demonstrate that different users may have a vastly different value for the service. We also analyse the reasons that some movie ratings may be more valuable than others and discuss the consequences for compensating users fairly.  
\end{abstract}

\section{INTRODUCTION}

Data, and the economy around it, are said to be driving the fourth industrial revolution. Interestingly -- the people -- whose data is what moves the new economy, have a rather passive role in it, as they are left outside the direct value flow that transforms raw data into huge monetary benefits. This is a consequence of the de facto understanding (or one may say misunderstanding) between people and companies, that the former get unpaid access to online services in exchange for unpaid access to their personal data. This is increasingly being challenged by various voices who call for the establishment of a new, renegotiated, relationship between users and services. Indeed, a variety of pathologies can be traced back to the way the data economy has been working so far. Some are direct and obvious, such as privacy risks for individuals, and market failures and dangers for the economy from the rise of data monopolies and oligopolies. Others are less obvious, and further reaching into the future, such as mass unemployment due to data-driven automation. 

It was estimated recently~\cite{posner} that, if automation due to artificial intelligence reaches maturity and fair remuneration algorithms are set in place, a family of four could earn up to \$20,000 per year from their data. 
The idea of micropayments, or providing small contributions to users in exchange for their presence on a platform or for accessing a service, is of course much older. In the pre-World Wide Web era, France developed a videotex online service called Minitel, that included micropayments as part of its design, but Jaron Lannier brought it to public attention in 2013, in his book "Who owns the future?"\cite{lanier}. In it, he argues that we have only underwent half of the Data Revolution, the part that compensates users with \emph{implicit} benefits, but not the part that also compensates them with \emph{explicit} monetary benefits. 

There have been a series of proposed approaches for how this compensation might materialise. The simplest, at least in theory, would be to assign a context-free value to data, a kind of dollar-per-bit measure. This has been proven to be very hard \cite{moody1999, king, alstyne, blackwell}. Indeed, since the value of data is strongly connected to its intended use, it becomes very difficult to argue about how to assign an a priori average value. For traditional currencies, we are able to have a context-free appreciation of their value for the simple reason that we have been using these currencies long enough to be able to do so. Although we clearly understand nowadays that one's browsing and mobility patterns, social network, or past purchases all have value, we are far from being able to appreciate how much this value is in terms of dollars or euros. The latter is further complicated by our inability to tell in advance, by how many parties, and how many times, a piece of data may be utilized. As an analogy, selling an individual's data, or rather renting it temporarily, is as difficult and risky as renting an infinitely fast vehicle, with no gas and maintenance costs, and without any prior restrictions with regard to mileage or the person driving it.

A second proposed method has been to compensate users for their privacy damage \cite{privacy1, privacy2}. Processing massive amounts of data can lead to privacy infringements, such as the leakage of habitual user behavior, their location or other personal identifiable information (PII). Users are thus seen as victims who must be compensated for their damage. 

Our approach is different, we consider users as active partners in the data value chain. Such a chain requires a business model, smart predictive algorithms for extracting useful information from raw data and online marketing for attracting and retaining users, among many others. The fundamental component of the value chain, however, is the user, and it is ultimately a matter of common sense that they should be rewarded in a fair manner, which may or may not exceed the perceived privacy-related damages.

In a Human-Centric Data Economy, when a transaction, or set of transactions, is converted, a proportion of the obtained revenue will be returned to the users. Defining the right amount to be returned to the users is difficult, as it depends on many market characteristics of a multilateral value chain, such as competition and user loyalty~\cite{courcubetis}. In this paper, we assume that the total amount of revenue to be redistributed to users is given, e.g., 5\% or 10\%, or any other number produced by the competition between services within a given sector. We thus turn our attention to the next important question, namely: given a fixed total amount to be distributed, how should this distribution be performed? How should different users be compensated based on the value of the data that they contribute to a specific service? 

One obvious answer would be to split the sum by the number of users on the platform. This may not be fair however, because it is unreasonable to assume that all users contribute equally to the service. In the case of a recommendation system, for example, some users may rate and view hundreds of items, while others may have a much sparser activity. Another example would be a traffic application. Users who travel regularly through a given area and constantly provide feed-back, may be more relevant to the platform than occasional passers-by. Even at an equal level of intensity, some movie scores, or location data may be more important for a recommender than others. For example a score for a blockbuster movie that just aired may be more important than a score for a well known (and voted) blockbuster movie. Similarly, speed information for a known-to-be-congested route is less useful to a navigation app than speed data for  an unknown alternative route with less traffic. 

Another important point to understand is that the Data Economy is not a zero-sum game. Paying users for data need not be seen as a measure that will reduce the revenue or profitability of online services. More and better data can lead to better services and thus more revenue and profit. It is thus not surprising that the vision of paying directly or through taxes for data has received positive comments from industry leaders such as Bill Gates \cite{bill}, Elon Musk \cite{theElon} and Mark Zuckerberg \cite{markFB}. Also, despite not happening today, there is a realistic path that can lead to wide adoption. It only requires that a small number of visionary companies start offering micro-payments to get a competitive advantage in terms of user retention, for others to follow, and the practice to get traction in the online services market. 

Our paper brings a threefold technical contribution towards the realisation of a Human-Centric Data Economy:

\begin{itemize}
\item We define data payoff \textit{fairness} in terms of accepted economic notions. Specifically, we use the Shapley value from collaborative game theory to define how to split a total payback among all the users that have provided data to a revenue-generating online service. We also sketch a Contribution-Reward framework for implementing such paybacks in practice.

\item We develop two algorithms that can provide efficient estimates of the Shapley value, which in its raw form does not scale for large data sets. The first one applies minibatch k-Means with $N$ d-dimensional points, for $k$ clusters and $t$ iterations directly on the definition of Shapley to reduce the complexity from $O(N!)$ to $O(N k d t + k!)$, where $k,d,t  \ll N$. The second is a $O(N^2)$ heuristic of local information that does not use the Shapley definition.
    
\item We apply the above algorithms to a movie scoring data set and study how different users may be in terms of their value for their service. We observe that some users may be as much as $\times 4$ more valuable than others.
    
    
\item Finally, we study how scoring behaviour impacts on user value. We find that what differentiates a ``good'' user from a ``bad'' one is that the good user tends to vote mainly for the most popular movies and is consistent with the movie popularity hierarchy, namely popular movies get high ratings and unpopular ones, low ratings. By contrast, users at the other end of the value distribution tend to vote and provide high scores to unpopular movies as well.
\end{itemize}

While human centricity and fairness need to reside at the core of the proposed framework, transparency must also be considered. Users need a method of verifying for themselves the amount of payment received and be able to connect their behaviour on an online platform to the obtained revenue, in a clearly interpretable manner.  We discuss how to do this via an accounting meta-data layer at the end of the article.  

\section{Background}

In the Background section, we introduce the Contribution-Reward framework and describe the setting for our two algorithms -- a recommender system generating recommendations for the users of an online movie streaming service. We also introduce the Shapley value, as the backbone of the first proposed algorithm and also an accepted point of reference in terms of fair credit distribution. The section ends with a toy example that shows how the exact Shapley value is computed in a simple two-user setting. 

\subsection{The Contribution-Reward Framework}

\begin{figure}
\centering
\includegraphics[width=0.5\textwidth]{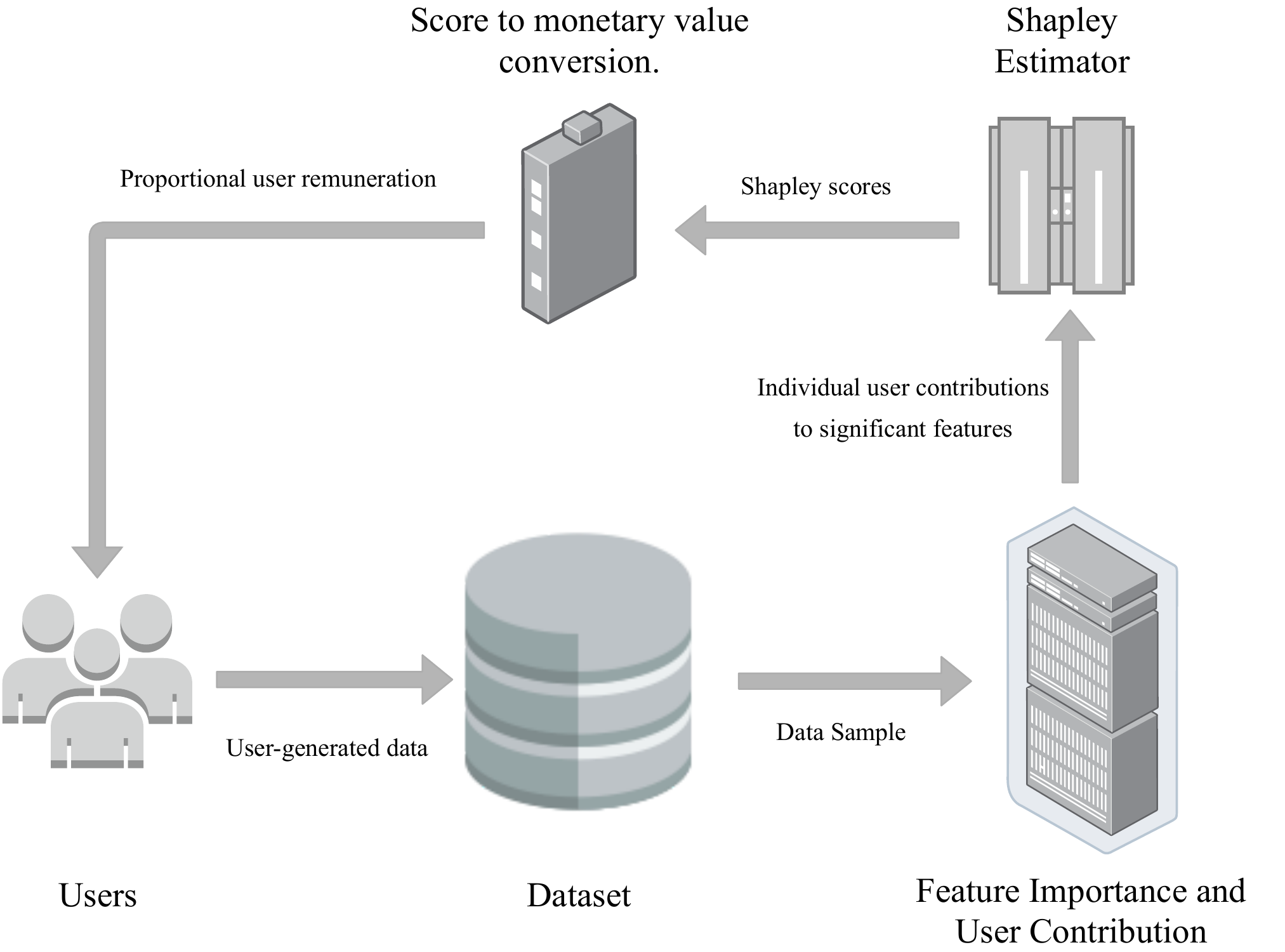}
\caption{The Contribution-Reward framework. In establishing the proportion of each user's contribution, one first determines important features or relevant measures that result in revenue. One such measure could be the number of useful recommendations obtained from a recommender system. After a measure or relevant feature has been identified, user contributions to it are computed and users are ordered in a hierarchy. Based on their place in the value hierarchy, they are remunerated.}
\label{shapley_scheme}
\end{figure}

The proposed Contribution-Reward framework, depicted in Fig. \ref{shapley_scheme}, aims to provide users with a remuneration scheme that is both fair and sustainable. With usage, the service gathers and stores relevant data. Based on the data type, a number of important features are chosen and the contribution of each user to the said features is estimated. Finally, the user receives a repayment from the platform, proportional to their estimated contribution. The difficulty of finding the said estimation resides in the need to first observe which are the relevant features that make a data set valuable. In the case of recommendation data, for example, it could be that users voting for the most popular movies, thus contributing to an already existing hierarchy, are considered more valuable, or on the contrary, users reviewing items which are initially unpopular could bring an element of novelty, and thus a higher overall value. On the other hand, if the accuracy and update rate of the data set are essential, for example in traffic applications, users with a higher contribution frequency may be considered of higher value. 

It thus becomes clear that user value depends on the structure of the data and its intended use. Due to this subjective nature, deriving a generally-applicable valuation framework is not a straightforward task. The question we raise here is two-fold: first, how can one determine a hierarchy of value for a given set of users and a particular use case (recommender systems) and second, how can one quantify the position of each user in the value hierarchy? An answer to the second question is of particular importance, since if one can assign scores to users, based on their contributions, one can then define a mapping from these scores to an actual financial amount. 

For the rest of this paper, the setting is thus fixed: a recommender system is trained on a training set, consisting of movie reviews, and then makes predictions on a separate test set. If the recommendations made to a user, on the test set, have an error below a fixed threshold, we consider that there is a high likelihood that the user will want to watch the recommended item. We refer to these recommendations as "clicks". As such, when we say: "user A has generated 5 clicks", this is to be interpreted as "the recommender has made 5 recommendations to user A, all with an error below our set threshold". Clicks here play the role of the \textit{important feature} in the Contribution-Reward framework. 

The most na\"{i}ve approach would be to train the recommender system by removing one user at a time, and counting the difference in the overall number of clicks obtained on the test set. There are a series of disadvantages to this: first, a data set may contain reviews from hundreds of thousands of users, quickly making the leave-one-out training computationally unfeasible. The second drawback resides in the fact that removing only a single user may not have the expected result on the recommender. Indeed, if the user provides a significant amount of novelty, it may happen that the system produces more clicks in the absence of the user than in their presence, leading to a negative assigned value. This would further be difficult to map to a monetary contribution. In the following sections, we present two alternative methods which avoid these shortcomings and can be implemented in a computationally-scalable way.  

After a theoretical discussion of the proposed algorithms, they are applied to a case study based on a subset of the MovieLens data set \cite{movielens}, depicting how groups of users can contribute differently to the overall performance of the recommender system, and thus hold different values with respect to the service.

\subsection{Introducing the Shapley Value}

Credit assignment in cooperative games has long been a central problem of cooperative game theory. To this end, Shapley \cite{shapley, shapley2} proposed that players should be rewarded in a manner proportional to their average marginal contribution to the payoff of any coalition they could join.

Let $\mathcal{N}$ be a set of $\mathcal{N}$ players and $\mathcal{S} \subset \mathcal{N}$ be a coalition with cost $v(S)$. The \textit{Shapley value} is a uniquely determined vector of the form $(\phi_1(v),..., \phi_n(v))$, where the element representing player $i$ is given by

\begin{equation}
\phi_i(v) = \frac{1}{N!} \sum_{\pi \in S_N} [v(S(\pi, i)) - v(S(\pi, i) \setminus i)],
\label{shapley_eq}
\end{equation}
where $\pi$ is a permutation representing the arrival order of set $\mathcal{N}$, while $S(\pi, i)$ represents the set of players that have arrived into the system before player $i$.

The Shapley value satisfies a series of important properties:

\begin{itemize}
    \item \textbf{Efficiency}: the total gain is completely distributed among the players
    \begin{equation}
        \sum_{i \in \mathcal{N}} \phi_i(v) = v(\mathcal{N})
    \end{equation}
    \item \textbf{Symmetry}: if $i$ and $j$ are two players who bring equal contributions, in the sense that, for every subset $S$ that contains neither $i$ nor $j$, $v(S \cup \{i\}) = v(S \cup \{j\})$, then their respective Shapley values are also equal, $\phi_i(v) = \phi_j(v)$.
    \item \textbf{Linearity}: if two coalition games, denoted $v$ and $w$ are combined, the resulting gain is the sum of the gains derived from each game separately
    \begin{equation}
        \phi_i(v + w) = \phi_i(v) + \phi_i(w)
    \end{equation}
    and also $\phi_i(\alpha \cdot v) = \alpha \cdot \phi_i(v)$, for any real $\alpha$.
    \item \textbf{Null player:} the Shapley value of a null player is zero. A player is considered null if they do not bring a contribution to any possible coalition. 
\end{itemize}

Unfortunately, the Shapley value has also been proven to be NP-hard for many domains \cite{papa, papa2, bachrach}. Since it takes into account all possible coalitions, for each user, the number of terms scales with $N!$, where $N$ represents the number of users, such that it quickly becomes computationally unfeasible.

In Ref.~\cite{nikos} the authors use Monte Carlo to approximate the Shapley value for computing the cost contribution of individual households to the peak hour traffic and costs of an Internet Service Provider (ISP). In that case, the relatively simpler structure of the problem made Monte Carlo an appropriate technique for approximating Shapley. Other recent works have presented approximation algorithms for Shapley for specific problems of lower complexity than recommendation \cite{shapapprox1, shapapprox2}. In the context of the current proposal, the inherent higher complexity of the considered value functions v() that may represent the workings of complex ML algorithms for things like recommendation, makes using Monte Carlo inaccurate according to our preliminary tests. Instead, we intend to use clustering to reduce the input size of the problem. Our approach will be to first group users in a number of clusters according to the similarity of their movie ratings (or trajectories in the case of mobility related applications), and then compute the Shapley value of each cluster instead of each user. By controlling the number of clusters we can make the computation as precise and complex as our computing resources allow. In Section III we will apply this method in the case of a movie recommendation data set, containing an amount of users for which the exact computation of Eq.~\ref{shapley_eq} is not possible.

\subsection{A Toy Example}

The purpose of this section is to provide a very simple example of computing the Shapley value for the case of two users. Consider an artificial data set, containing movie recommendations from two different users. The predictions made by the recommender are the net contribution in our case. We start from the marginal contributions $V(\{1\}) = 12$ and $V(\{2\}) = 15$, which would mean that the presence of user 1 alone in the data set, causes the recommender to produce 12 useful movie recommendations, and the presence of user 2 alone results in 15 useful recommendations. Let us further assume that the presence of both users, simultaneously, increases the number of recommendations to 28, hence $V(\{1, 2\}) = 28$. Since it is clear that the two contributions are not equal, the following question arises: how can we find a factor, proportional to the user contributions, that would be useful in determining a fair repayment ?

For this, we compute the Shapley value, as defined in Eq.~\ref{shapley_eq}. We first note that there are two possible orders of arrival of the users, with equally likely probabilities of occurrence: $[1, 2]$ or $[2, 1]$. In the first situation, user 1 comes first, bringing a contribution of $V(\{1\}) = 12$, followed by user 2, who increases the overall useful contribution to $V(\{1, 2\}) = 28$, thus, the net contributions of the two users, for this particular order of arrival are

\begin{equation*}
\begin{split}
\phi_{[1,2]}(1) &= V(\{1\}) = 12, \\
\phi_{[1,2]}(2) &= V(\{1, 2\}) - V(\{1\}) = 28 - 12 = 16. 
\end{split}
\end{equation*}

In the second case, the order of arrival is reversed, namely $[2, 1]$. User 2 arrives first, bringing a contribution $V(\{2\}) = 15$, followed by user 1 who increases the net contribution to $V(\{1, 2\}) = 28$, thus

\begin{equation*}
\begin{split}
\phi_{[2,1]}(1) &= V(\{1, 2\}) - V(\{2\}) = 28 - 15 = 13, \\
\phi_{[2,1]}(2) &= V(\{2\}) = 15.
\end{split}
\end{equation*}

Since the two cases are equally probable, the Shapley value is the sum of the two marginal contributions multiplied by the probability of the order of arrival (or of the factorial of the number of users)

\begin{equation*}
\begin{split}
\phi(1) &= \frac{1}{2!} \cdot (\phi_{[1,2]}(1) + \phi_{[2,1]}(1)) = 12.5, \\
\phi(2) &= \frac{1}{2!} \cdot (\phi_{[1,2]}(2) + \phi_{[2,1]}(2)) = 15.5. \\
\end{split}
\end{equation*}

We thus see that the average marginal contribution of user 2 is higher than the one of user 1, and based on this factor, we can devise a scheme of fair remuneration, to be presented as part of the Contribution-Reward framework.

It is also worth noting that, in order to compute the Shapley terms, we need to compute $N!$ marginal contributions, which leads to scaling problems if an exact computation of the Shapley value is required. The methods presented in this paper provide ways of overcoming this obstacle through various approximate approaches.

\section{User Value Estimation}

In this section we first present and then apply the two proposed algorithms. We then discuss the results obtained on a subset of the MovieLens data set. At the end of this section and in the next, we will interpret user behaviour and try to understand what is the relationship between votes given, number of votes, types of movies reviewed and the assigned user value. 

\subsection{Approximate Shapley Value Estimation (ASVE)}

The first user contribution estimation method is based on the Shapley value, described in the previous sections. The algorithm is constructed around a recommender system framework and we choose our value of interest to be the number of "clicks" or useful recommendations that the model provides, for a given data set. The pseudocode for this algorithm is provided as Algorithm \ref{shapley_algo}.

The input data consists of user identifiers, product identifiers (movie categorical IDs) and votes. From Eq.~\ref{shapley_eq} we observe that computing the Shapley value of each user in the data set directly is unfeasible, as the computational complexity is of order $N!$ where $N$ (the number of users in the data set) is typically extremely large (in the order of hundreds of thousands or even millions for services like YouTube and NetFlix). Clearly such an approach does not scale.  

There are two plausible directions one can pursue in order to avoid the complexity barrier. The first is to attempt to estimate the Shapley value using Monte Carlo sampling, as done in Ref.~\cite{nikos}. The second, which we shall employ here, takes advantage of the similarities between user behavioural patterns. In general, when treating data related to user preference, consumers tend to cluster into a limited number of similar groups. 

\begin{algorithm}
\caption{Approximate Shapley Value Estimation}\label{shapley_algo}
\begin{algorithmic}[1]
\Procedure{ASVE}{}
\State \text{\footnotesize{\# Cluster the users based on behaviour.}}
\State $\text{clusteredData, clusterLabels} \gets \text{Cluster} \text{(inputData)}$
\State $\text{clusterCoalitions} \gets \text{Compute} \text{(clusterLabels)}$
\State \text{\footnotesize{\# Compute the total number of clicks.}}
\State $ K \gets \text{TrainPredict(inputDataset)} $
\State \text{\footnotesize{\# Compute marginal contribution of every possible coalition.}}
\State \text{\footnotesize{\# trainData and predData contain only clusters from coalition.}}
\For {S in clusterCoalitions}
    \State $V_{S} \gets \text{TrainPredict(trainData, predData)}$
\EndFor
\State \text{\footnotesize{\# Create all possible permutations of complete cluster coalitions.}}
\State \text{\footnotesize{\# and compute the marginal contribution of each cluster.}}
\State $\text{perm} \gets \text{Permute}\text{(clusterLabels)}$
\For {$\text{$\pi$} \text{ in } \text{perm}$}  
    \For {$\text{i} \text{ in } \text{clusterLabels}$}  
        \State $V_{i} \gets V(\pi) - V(\pi - \{i\})$
        \State \text{margContrib[i]} += $V_{i}$
    \EndFor
\EndFor
\For {$\text{i} \text{ in } \text{clusterLabels}$} 
    \State $\phi_{i} = \frac{1}{len(clusterLabels)!}*\text{sum(margContrib[i])}$ 
\EndFor
\State \text{\footnotesize{\# Compute the value of every user.}}
\For {$\text{userId} \text{ in } \text{clusteredData}$} 
    \State $\tilde{\phi}_{1}\text{(userId)}$ = $(\sum_{k} \frac{\phi_{k}}{1 + \text{EuclideanDist}\text{(user, centroid(k))}})$ 
    \State \text{userValSum} += $\tilde{\phi}_{1}\text{(userId)}$
\EndFor
\State \text{\footnotesize{\# Normalize user values to respect efficiency condition.}}
\For {$\text{userId} \text{ in } \text{clusteredData}$} 
    \State $\phi_{1}\text{(userId)}$ = $\tilde{\phi}_{1}\text{(userId)} \cdot \frac{K}{\text{userValSum}}$
\EndFor
\EndProcedure
\end{algorithmic}
\end{algorithm}

\begin{algorithm}
\caption{Neighbourhood Similarity Value Estimation}\label{method3}
\begin{algorithmic}[1]
\Procedure{NSVE}{}
\State $ \text{Rec} = 0 $
\State \text{\footnotesize{\# Train model once to obtain the total number of clicks.}}
\State $ K \gets \text{TrainPredict(inputDataset)} $
\State \text{\footnotesize{\# For all neighbors in a neighbourhood, compute the user's}}
\State \text{\footnotesize{\# predictive potential}}
\For {$\text{userId} \text{ in } \text{inputDataset}$} 
\State $\text{neighbors} \gets \text{FindNeighbors}\text{(inputDataset, userId)}$ 
\For {$\text{id} \text{ in } \text{neighbors}$}
\State $ \text{newItems} \gets \text{IntersectLeft(userId, id)} $
\State $ \text{Rec} += 1 $
\EndFor
\State $ \text{RecList}\text{.append}\text{(Rec)}$
\State $ \text{Rec} \gets 0 $
\EndFor
\State \text{\footnotesize{\# Scale the predictive potential by the total}}
\State \text{\footnotesize{\# number of clicks divided by the sum of the predictive}}
\State \text{\footnotesize{\# potentials, to respect the efficiency condition.}}
\For {$\text{userRec} \text{ in } \text{RecList}$} 
\State $ \phi_{3}\text{(userId)} \gets \frac{\text{K*userRec}}{\sum_{j}\text{userRec}_{j}} $
\EndFor
\EndProcedure
\end{algorithmic}
\end{algorithm}

For example there may be individuals with a stronger preference for action movies as opposed to romantic ones. Exploiting such relationships allows us to greatly simplify the estimate of a user's contribution to the overall data set.

The algorithm starts by clustering the input data, prior to which, a dimensionality reduction method (such as PCA) is applied. The next step is the calculation of all possible coalitions of clusters. For example, for three clusters, these are

\begin{equation*}
   \{1\}, \{2\}, \{3\}, \{1, 2\}, \{1, 3\}, \{2, 3\}, \{1, 2, 3\},
\end{equation*}
where $\{1\}$ represents a data set where only the users corresponding to the first clusters are present. The recommender system is then trained on the filtered data sets, corresponding only to the clusters in the above coalitions. A train-test split is performed prior to this, but omitted in line 8 of the pseudocode for clarity. During the prediction phase, the model makes recommendations to users and also provides an error estimate for each recommendation. We assert that, for errors under a certain threshold, the recommendations can be valuable, and we refer to these recommendations as "clicks" for the rest of this paper, in analogy to advertising recommendations where, if the user is interested in an advert, they will click the banner. 

After obtaining the number of clicks for every one of the possible subsets of clusters, one must compute the marginal contribution of the cluster. This takes into account every possible complete coalition, where the order of arrival matters. For example, in our three user case, all possible complete coalitions are: $\{1, 2, 3\}, \{2, 1, 3\}$ and $\{3, 2, 1\}$. The relevant code is between lines 11 - 19. 

Finally, when the Shapley value (computed in line 19) for each cluster is known, we can proceed to determine the value of each individual user. For this, the centroid of the cluster is labeled with the corresponding Shapley value of the cluster. Thus, the value of each user is equal to the sum of all individual cluster values divided by the Euclidean distance between the point, representing the user, and the respective cluster centroid, with one added for stability. The assigned values, from both proposed algorithms, are then scaled with the total number of clicks produced by the recommender, on the complete data set, to ensure that the efficiency condition is met. In this manner, to every point on a projected two-dimensional surface, representing the space of all users, we can assign an approximate Shapley value.

\begin{figure*}
  \includegraphics[width=1.01\textwidth,height=4.5cm]{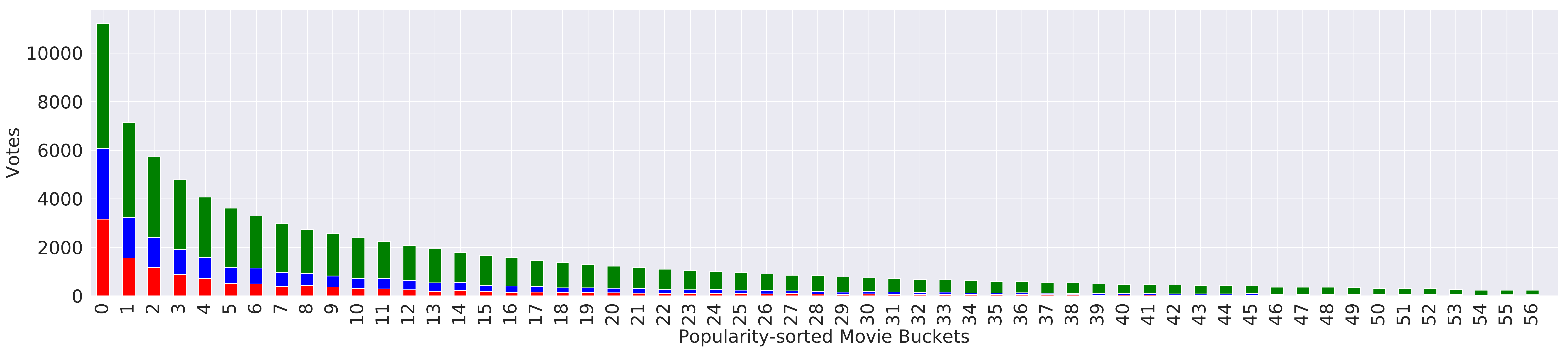}
  \caption{User vote distributions based on movie popularity for the ASVE method. Votes from the most valuable users are shown in red, those from average users in blue and votes from the least valuable users are shown in green. For clarity, the movies have been grouped into ordered buckets, with popularity decreasing in the positive direction of the horizontal axis.}
  \label{pop_voting_sve}
\end{figure*}

\begin{figure*}
  \includegraphics[width=\textwidth,height=4.5cm]{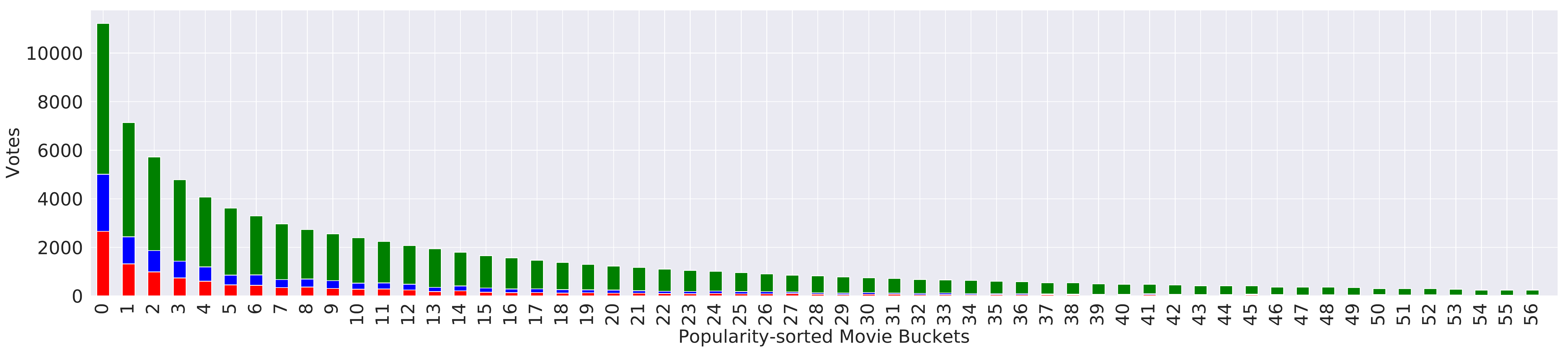}
  \caption{User vote distributions based on movie popularity for the NSVE method. Votes from the most valuable users are shown in red, those from average users in blue and votes from the least valuable users are shown in green. For clarity, the movies have been grouped into ordered buckets, with popularity decreasing in the positive direction of the horizontal axis.}
  \label{pop_voting_nve}
\end{figure*}

We have applied the ASVE algorithm to a subset of the MovieLens data set, containing 92,394 total ratings on 4,180 movies from 610 users, with the intent on understanding if the value hierarchy provided by the method can be intuitively understood. First, the users were separated into three classes, the best users (representing the ones with the highest scores given by the ASVE method), average users and bad users. We considered whether or not there is a relationship between movie popularity (based on the overall number of ratings a movie has received) and the vote distribution of the three classes of users. 

In Fig.~\ref{pop_voting_sve}, the user vote distributions are presented, for the three classes of users. The movies are grouped into equal-sized ordered buckets, with their popularity decreasing in the positive direction of the horizontal axis. We observe that users which the ASVE method considers as being the most valuable, tend to vote mostly on popular movies, whereas users considered to have low value, have a more widespread distribution of votes. This reinforces our initial assumption that the ASVE method gives high scores to users who agree with a predetermined hierarchy, not necessarily to users with a novel contribution. 

One remark that must be made is that, by design, the application of ASVE requires one prior check. When removing a cluster of users, it is entirely possible that one or more clusters provide a negative contribution. It is therefore essential that one checks the values assigned to the clusters for non-negativity, prior to applying the method to individual users.

\subsection{Neighbourhood Similarity Value Estimation (NSVE)}

The second approach relies on first reducing the dimensionality of the data set (similar to ASVE) and creating a neighbourhood around each user. Based on the predictive capability of the user on its neighbours, one can approximate the user's importance. 

The algorithm works as follows: first, the recommender system is trained on the complete data set and a total number of clicks (similar to the previous method) is obtained (line 4). For each user in the data set, we then construct a neighbourhood and observe the number of items that the central user has rated, and the neighbours have not. These elements act as an estimator of the user's "predictive potential". For example, if user A has seen and rated movies M1, M2 and M5 and user B, which is in user A's neighbourhood, has only seen movie M1 (out of the list that user A has rated overall), then A could potentially recommend movies M2 and M5 to B, so we say that "A has a predictive potential of 2 (movies) over B). This predictive potential is then used, in line 18, to proportionately assign clicks (from the total number K) to each user, based on their respective predictive potentials. 

One must keep in mind that the size of the neighbourhood is a hyperparameter that must be determined through trial and error. Furthermore, there may be other distance metrics, other than the Euclidean distance, chosen here, that could potentially offer a better estimate, however, for simplicity, we restrict ourselves to Euclidean distances in the present paper. 

\begin{figure}
  \includegraphics[width=0.5\textwidth]{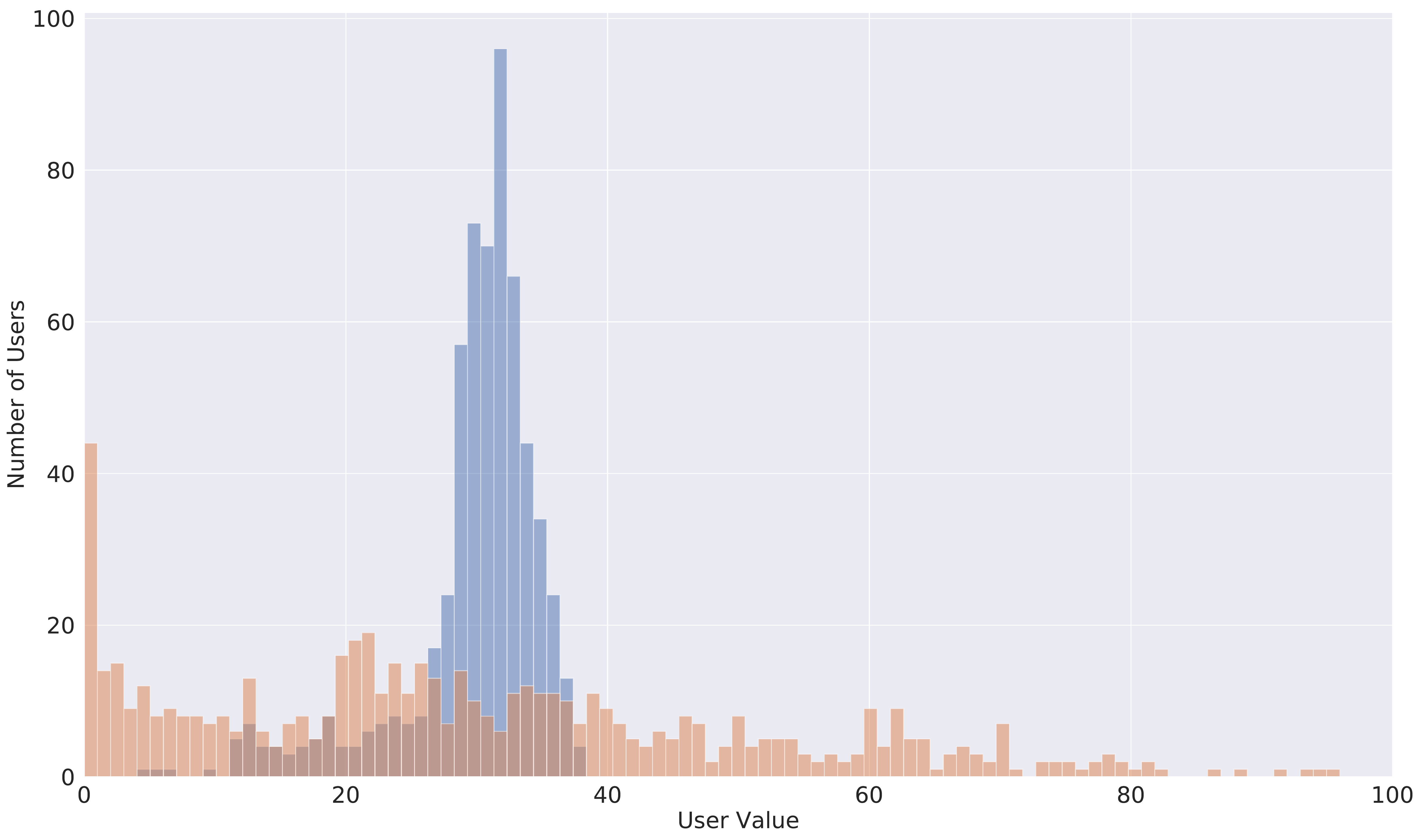}
  \caption{The ASVE distribution, shown in blue, and the unscaled NSVE distribution, shown in orange. It is important to note that, even though the variance of the two distributions is different, the user value hierarchy is maintained to a large degree.}
  \label{unscaled_dist}
\end{figure}

\begin{figure}
  \includegraphics[width=0.5\textwidth]{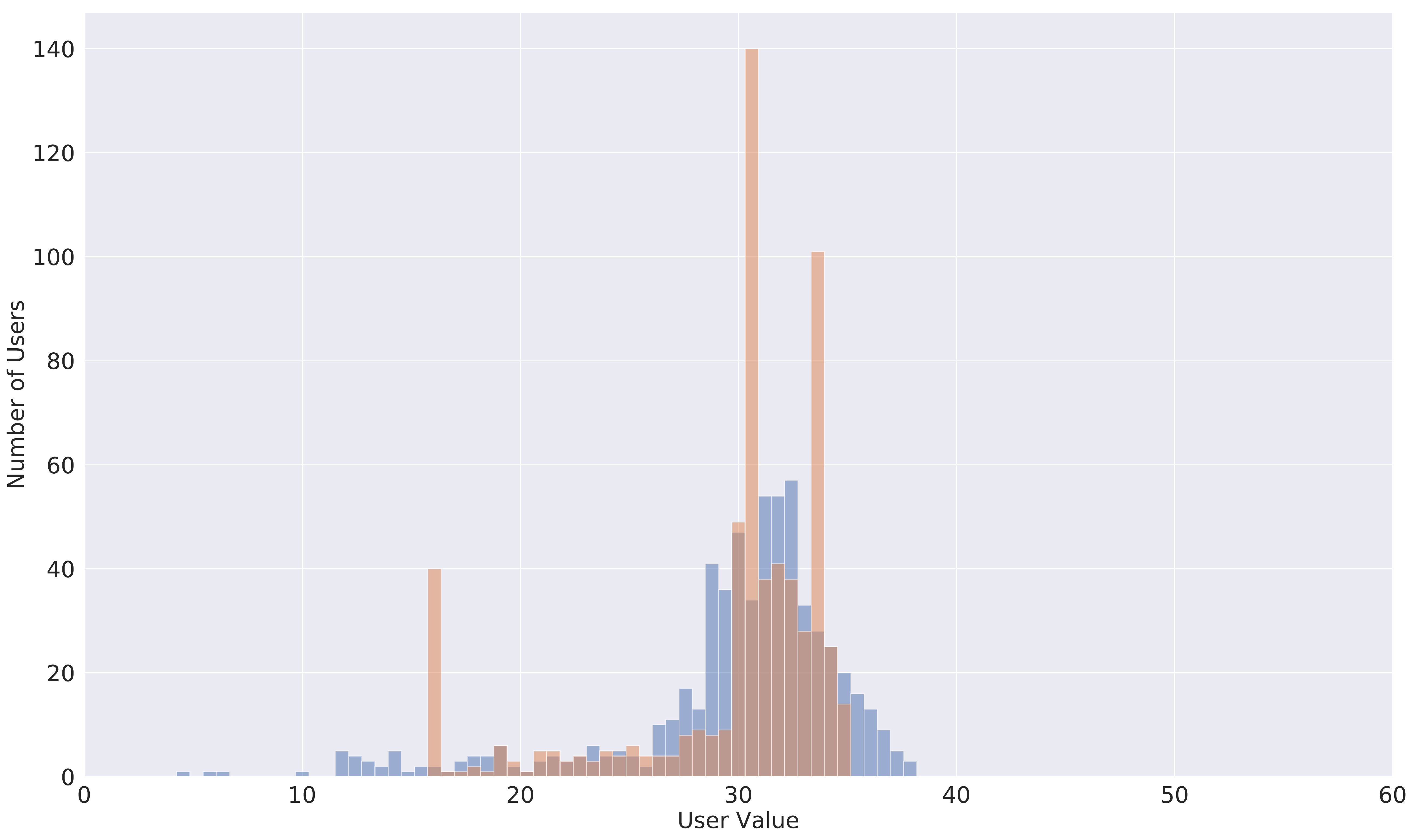}
  \caption{The ASVE distribution, in blue, and scaled NSVE distribution, in orange, showing artefacts due to the approximate mapping between the two.}
  \label{scaled_dist}
\end{figure}

In order to compare the two methods, a proper scaling must be found for the two corresponding distributions. The unscaled distributions, shown in Fig.~\ref{unscaled_dist} are correlated, with a correlation distance of 0.29 (0 representing a perfect correlation) for the MovieLens data set. We were thus able to learn an approximate mapping, that would project the second distribution to the domain of the first, making a direct comparison possible, as depicted in Fig.~\ref{scaled_dist}.

Due to the nature of this mapping being approximate, one sees artefacts, such as the unusually tall peaks of the projected NSVE distribution, compared to its ASVE counterpart. While this is an issue that needs to be considered when comparing method predictions, the mapping process would not be necessary in a real-world implementation, as it's sole purpose is to aid the comparison, and would be irrelevant for the purpose of credit assignment. 

Having the two distributions on the same scale, we continued our investigation by asking whether or not the second method produces the same distribution of user votes, based on movie popularity. As seen in Fig~\ref{pop_voting_nve}, both methods agree on the fact that users considered valuable tend to vote for popular movies. 

It is instructive, at this point, to also see how users voted, based on movie popularity, that is, is it the case that more popular movies got higher ratings than less popular ones? Also, how do the three classes of users rate movies of various levels of popularity? 

\begin{figure*}
  \includegraphics[width=1.01\textwidth,height=4.5cm]{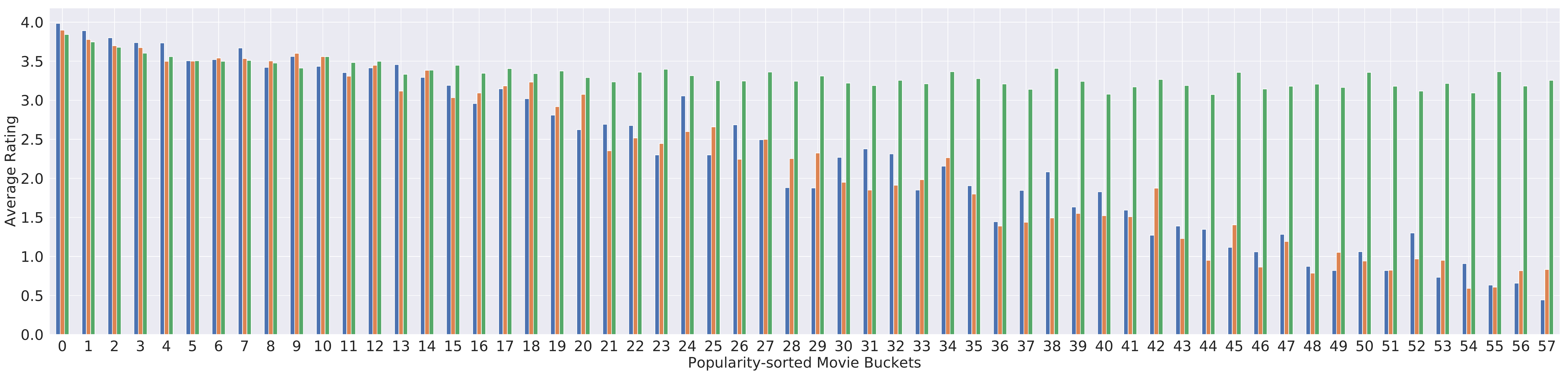}
  \caption{Movie ratings given by the three user classes, sorted by movie popularity. The average ratings from the best users (over both methods) are depicted in blue, those from users with an intermediate value, in orange and those from the least valuable users, in green.}
  \label{ratings_by_class}
\end{figure*}

In Fig.~\ref{ratings_by_class}, we present the vote distribution by user category. Rather surprisingly, there is somewhat little difference between users considered \textit{very valuable} and those considered \textit{average}, as both classes seem to give high scores to very popular movies and low scores to unpopular ones. Less valuable users (according to both methods) seem to offer high scores to a wide array of movies, going against the established popularity hierarchy. This further serves to confirm that users reinforcing an existing order are rated higher.

One point to notice is that, in the case of the NSVE, by design a user cannot have a negative assigned value, thus no prior checks are required, as in the case of ASVE. A potential problem here is the reliance of the method on the neighbourhood radius hyperparameter, which strongly depends on the data set. Furthermore, users outside a given neighbourhood have zero contribution to the final rating of the neighbourhood source. This is in stark contrast to the smooth value assignment of ASVE, and does give rise to a series of artefacts, one of which being that there are always (or at least in all practical situations) users with an assigned zero value, which raises the difficulty of mapping this to an actual financial amount.

\begin{figure*}
  \includegraphics[width=\textwidth,height=4.5cm]{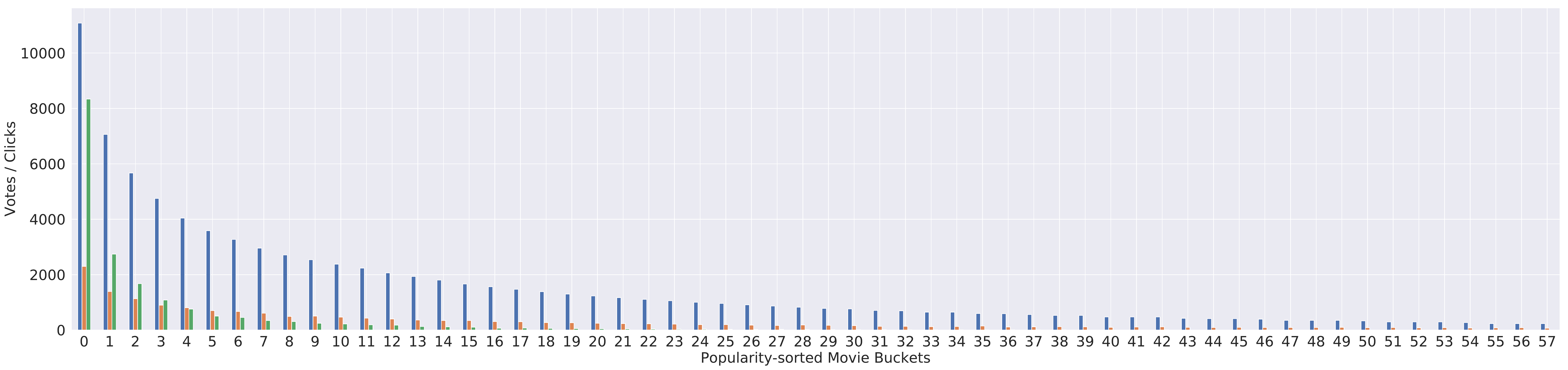}
  \caption{Movie votes and number of clicks generated by the corresponding bucket, according to the two credit-assignment methods. The overall number of votes is shown in blue, the number of clicks that movies in the corresponding bucket generate are shown in orange, for ASVE and green for NSVE.}
  \label{ratings_by_pop}
\end{figure*}

A second artefact can be observed once we ask the question: how many clicks do movies generate (over all users in a fixed test set) based on their popularity? As can be seen in Fig.~\ref{ratings_by_pop}, the Neighbourhood-based method is very sensitive to the relative distance between users. As movies become less popular and users who vote for them get spread over larger distances, the amount of clicks generated (which in the case of NSVE represent movies that the central user can recommend to its immediate neighbours) quickly falls to zero. Thus, NSVE strongly favors users who vote for popular movies, where users who favor them cluster together.

From a computational standpoint, the NSVE algorithm only requires one training of the recommender, in order to obtain the total number of clicks. The neighbourhood comparison operations are linear. Another advantage of this method is that, if applied to a production environment, the model would continuously train as new user data becomes available, and only be required to perform occasional inference, to update the total number of clicks. 

Another potential impediment to both methods is their sensitivity to badly-labelled data. If, for example, in our movie data set, the tags describing the movie genre are not consistent or worse, misleading, this will affect the distances between users and result in both methods providing inaccurate value estimates. Therefore, data cleaning and a careful analysis of the data set are important, before the algorithms are employed. 

\section{Discussion}

In the previous section, we have presented two credit-assignment methods that could become part of an online service, and transform users into active participants in the information value flow, from its raw form to the final generated revenue. As the main producers of raw data, fairness requires that users share part of the resulting revenue, in a manner proportional to the significance of their data. 

We have proposed two different methods that are in agreement on a few key points. The first method is based on the commonly-used Shapley value, which is approximated by being computed at cluster level. The users are then scored based on their relative distance from the cluster centers. The second method exploits user similarity, creating a neighbourhood around each user, where users with similar preferences may be found. It is thus assumed that the items the central user can recommend are close to the interests of his neighbours, such that they become potential \textit{clicks}. This is a very crude, but also very effective method for performing distance-based recommendations. 

We have seen that the two algorithms provide consistent predictions on key aspects such as the distribution of user votes based on their attributed value, the score given by users based on the same attributed value and the number of \textit{clicks} movies produce, based on their popularity. These come to confirm our initial hypothesis, based on intuition, that users receiving high scores vote in line with the item's popularity level. There are, however, points where the two methods do not agree. The NSVE is more sensitive to distances between users in the projected plane. In areas where the users are sparsely distributed, NSVE tends to assign zero values to them. This is not the case for the ASVE algorithm, as belonging to a cluster ensures a non-zero contribution to the final outcome. 

Also as a result of the sharp cut-off at the neighborhood's boundary, the scores assigned to users by the Shapley-based method change in a much smoother and more continuous manner from one region to another. This reflects itself in a much narrower distribution (as seen in Fig.~\ref{unscaled_dist}). In contrast, the user score distribution of the NSVE method has a very high variance, assigning very small scores to users who are far apart and very high scores to the ones in densely-populated regions. 

One issue not yet discussed is how should new users be treated, with respect to their overall value. As their contribution is minimal, a valid argument would be to start by assigning them relatively small scores (hence small proportional payments). This is indeed the case for both methods. Considering ASVE, a new user would likely be far from the center of any cluster, and, as the proportional payment decreases linearly with the distance from the cluster center, would get a small reward. For NSVE the situation is rather similar, a new user would either be in a sparsely-populated area of the projected space or in a densely-populated area but having very few possible recommendations for their neighbours. In either case, they would receive a small score.

Finally, what should be the impact of low votes? This question is of particular importance if either of the two proposed algorithms is to be implemented in a realistic setting. Observing Fig. 6, we see that valuable users give high votes to popular films and low votes for unpopular ones. If this rule were to hold for a vast array of data sets, users could attempt to \textit{game the system}, by voting based on popularity and hence obtaining a high reward. This is also a problem that stems from the objective of the two methods, namely maximizing the number of clicks for each item and not necessarily considering the element of novelty brought into the system by each individual.

\section{Conclusion}

The current business model for online services is reaching the limits of sustainability. The exchange of free services for free data raises concerns both in terms of safety, with users allowing highly sensitive private information to be used as a currency, and also in terms of fairness, as an ever increasing number of voices call for an evaluation of the real worth of an individual's online impact. 

In the present paper, we have proposed a framework to determine the value of every user for an online service, based on their contribution to a quantity or metric of interest. The proposed framework must be fair, such that users are rewarded proportionally to their contribution and transparent, the payments need to be interpretable and linked to user behaviour on the platform. 

By restricting ourselves to the specific case of a recommender system, we aim to gain an understanding of what type of behaviour increases the overall worth of an individual to the service. Once the profile of valuable users has been studied on a series of particular cases, our goal is to generalize and establish value in a broader sense. A second direction of future research is the creation of a transparent accounting meta-data layer that users can access to make sure that they have been fairly compensated, and do that in a manner that does not harm their privacy by leaking data. 

As our goal is to ultimately expand our analysis to a broad spectrum of domains, the proposed Contribution-Reward framework aims to be as wide-reaching as possible. There are no mentions regarding the types of data or its use cases. The framework simply states that, once a set of important features or relevant metrics can be identified, based on data type and use case, a value hierarchy for users can be established. The selection of the metrics or features is done in terms of their impact on the overall revenue produced by the service, and this selection may not be a trivial procedure. Once this has been achieved, however, the second step of the framework is the ranking of users, and this can be done with algorithms similar to the ones proposed in our case study. The obstacles reside not only in determining the value of an individual, but also in ensuring the safety of their data and fairness of their remuneration. Once every user has been assigned a score, the final step of the framework consists in mapping the scores to actual financial quantities. This is certainly not a static map, as it may depend on a series of continuously-varying parameters, such as revenue produced, amount of revenue to be distributed among users and possibly even exchange rates between different currencies. In this paper we cover the second problem, which is concerned with ranking.

Following this line of reasoning, we have presented two algorithms for assigning proportional value to users, for a specific task. We then saw that these algorithms do agree on a number of points, despite the fact that both rely on approximations. We also introduced the exact method for determining proportional contributions, the Shapley value, which in its raw form does not scale to large coalitions, but can serve as a reference or the basis for approximate methods.

The predictions of the two algorithms are in line with common-sense assessments of the value of a particular user, for example with respect to their votes for the most popular, and hence revenue-producing movies. While this does offer further validation of the two methods, their limitations have also been highlighted, and these stem from the fact that both are, to different degrees, estimates of an exact but uncomputable estimate. 

As this field, which may aptly be named \textit{Data Economics} is still in its infancy, the number of research directions is rather considerable. One such path would be to treat other particular cases, such as traffic data and the case of a traffic application, and understand the reasons behind the difference in value for various types of users. One can then hope that, having understood a number of such cases, they will provide some intuition into possible generalizations. A different research direction would be to address the other two components of the reward framework, namely feature or metric identification and score-to-currency mapping.

An important aspect of our future work is designing and implementing a layer of transparency in the Contribution-Reward framework, with the explicit aim of allowing users to verify the manner in which their repayments are distributed and connected to their behaviour. Transparency is a fundamental characteristic, as it will both increase user confidence in regard to participating on a platform which implements the proposed framework; and growing confidence will, in turn, boost user presence and, hence, generated revenue. 

The purpose of the present paper has been to first pose a series of essential questions in Data Economics, especially related to fair credit assignment, and to show that, for a particular use case, a transparent algorithmic solution can be found, based on a commonly-accepted economic method, the Shapley value. It is our hope that, as our data economy reaches maturity, the remaining open problems in the field will soon be addressed.  


{}


\begin{thebibliography}{99}
\balance
\bibitem{posner} Posner E. A. and Weyl E. G., Radical Markets Uprooting Capitalism and Democracy for a Just Society. Princeton University Press, 2018, ISBN 9780691177502.
\bibitem{lanier} Lanier J., Who owns the future?. Simon \& Schuster, 2013, ISBN 9781451654967.
\bibitem{moody1999} Moody D. L. and Walsh P., Measuring the value of information-an asset valuation approach. ECIS, (496 -- 512), 1999.
\bibitem{king} King K., A case study in the valuation of a database. Journal of Database Marketing \& Customer Strategy, \textbf{14}, 2, (110 -- 119), 2007.  \textit{https://doi.org/10.1057/palgrave.dbm.3250041}
\bibitem{alstyne} Alstyne, M. W., A Proposal for Valuing Information and Instrumental Goods. Proceedings of the 20th International Conference on Information Systems, \textbf{20}, (328 -- 345), 1999. 
\bibitem{courcubetis} Gyarmati L., Laoutaris N., Sdrolias K., Rodriguez P. and Courcoubetis C., From advertising profits to bandwidth prices-A quantitative methodology for negotiating premium peering. NetEcon'14, 2014. \textit{"arXiv:1404.4208v4} 
\bibitem{bill} Delaney K. J., The robot that takes your job should pay taxes, says Bill Gates. 2017. Retrieved from: \textit{https://qz.com/911968/bill-gates-the-robot-that-takes-your-job-should-pay-taxes/}
\bibitem{theElon} Thomas L., Universal basic income debate sharpens as observers grasp for solutions to inequality. 2017. Retrieved from: \textit{https://www.cnbc.com/2017/03/25/universal-basic-income-debate-sharpens.html}
\bibitem{markFB} Gillespie P., Mark Zuckerberg supports universal basic income. What is it?. 2017. Retrieved from: \textit{https://money.cnn.com/2017/05/26/news/economy/mark-zuckerberg-universal-basic-income/index.html}
\bibitem{shapapprox1} Cabello S. and Chan T. M., Computing Shapley values in the plane. 2018.
\textit{arXiv:1804.03894}
\bibitem{shapapprox2} Zhao K., Mahboobi S. H. and Bagheri S. R., Shapley Value Methods for Attribution Modeling in Online Advertising. 2018. \textit{arXiv:1804.05327}
\bibitem{blackwell} Blackwell D., Equivalent Comparisons of Experiments. Annals of Mathematical Statistics, \textbf{24}, (265 -- 272), 1953.
\bibitem{privacy1} Carrascal J. P., Riederer C., Erramilli V., Cherubini  M. and Oliveira R., Your browsing behavior for a big mac: economics of personal information online. Proceedings of the 22nd international conference on World Wide Web, (189 -- 200), 2013.
\bibitem{privacy2} Acquisti A., Taylor C. R. and Wagman L., The Economics of Privacy. Journal of Economic Literature, \textbf{54}, 2, 2016.
\bibitem{movielens} Harper F. M. and Konstan J. A., The MovieLens Datasets: History and Context. ACM Transactions on Interactive Intelligent Systems, \textbf{5}, 4, Article No. 19, 2016.
\bibitem{shapley} Shapley L. S., A Value for n-Person Games. Annals of Mathematics Study, \textbf{28}, (307 -- 317), 1953.
\bibitem{shapley2} Winter E., The Shapley Value. Published in The Handbook of GameTheory, North-Holland, 2002, ISBN 9780444894281.
\bibitem{papa} Papadimitriou C. H., Computatational Complexity, Addison-Wesley, 1994, ISBN 9780201530827.
\bibitem{papa2} Deng X. and Papadimitriou C. H., On the Complexity of Cooperative Solution Concepts. Mathematics of Operations Research, \textbf{19}, 2, (257 -- 266), 1994. 
\bibitem{bachrach} Bachrach Y., Elkind E., Meir R., Pasechnik D., Zuckerman M. Rothe J. and Rosenschein J. S., The Cost of Stability in Coalitional Games. Proceedings of SAGT, (112 -- 134), 2009.
\bibitem{nikos} Stanojevic R., Laoutaris N. and Rodriguez P., On Economic Heavy Hitters: Shapley value analysis of 95th-percentile pricing. IMC'10 Melbourne, Australia, November 1-3, 2010.  
\end{thebibliography}
\end{document}